\documentclass[aps,preprint]{revtex4}%
\usepackage{amsfonts}
\usepackage{amsmath}
\usepackage{amssymb}
\usepackage{graphicx}%
\providecommand{\U}[1]{\protect\rule{.1in}{.1in}}
\newtheorem{theorem}{Theorem}

\newtheorem{remark}[theorem]{Remark}

\begin{document}
\preprint{ }
\title[Short title for running header]{Quaternionic (super)twistors extensions and general superspaces}
\author{Diego Julio Cirilo-Lombardo}
\affiliation{Universidad de Buenos Aires, Consejo Nacional de Investigaciones Cientificas y
Tecnicas (CONICET), National Institute of Plasma Physics (INFIP), Facultad de
Ciencias Exactas y Naturales, Ciudad Universitaria Buenos Aires 1428, Argentina}
\affiliation{Bogoliubov Laboratory of Theoretical Physics, Joint Institute for Nuclear
Research, 141980 Dubna, Russia}
\author{Victor N. Pervushin}
\affiliation{Bogoliubov Laboratory of Theoretical Physics, Joint Institute for Nuclear
Research, 141980 Dubna, Russia}

\begin{abstract}
In a attempt to treat a supergravity as a tensor representation, the
$4$-dimensional $N$-extended quaternionic superspaces are constructed from the
(diffeomorphyc)graded extension of the ordinary Penrose-twistor formulation,
performed in a previous work of the authors\cite{diegoIJGMMP}, with $N=p+k.$
These quaternionic superspaces have $4+k\left(  N-k\right)  $
even-quaternionic coordinates and $4N$ odd-quaternionic coordinates where each
coordinate is a quaternion composed by four $\mathbb{C}$fields (bosons and
fermions respectively). The fields content as the dimensionality (even and odd
sectors) of these superspaces are given and exemplified by selected physical
cases. In this case the number of fields of the supergravity is determined by
the number of components of the tensor representation of the 4-dimensional
N-extended quaternionic superspaces. The role of tensorial central charges for
any $N$ \textit{even }$USp\left(  N\right)  =Sp\left(  N,\mathbb{H}%
_{\mathbb{C}}\right)  \cap U\left(  N,\mathbb{H}_{\mathbb{C}}\right)  $ is
elucidated from this theoretical context.

\end{abstract}
\volumeyear{year}
\volumenumber{number}
\issuenumber{number}
\eid{identifier}
\date[Date text]{date}
\received[Received text]{date}

\revised[Revised text]{date}

\accepted[Accepted text]{date}

\published[Published text]{date}

\startpage{101}
\endpage{102}
\maketitle
\tableofcontents

\section{Introduction}

In theoretical physics from long ago, there are an increment of the use of
modern mathematical methods to treat several problems of diverse degrees of
complexity\cite{olg}.. From the hydrodynamics and the mechanics of the
continous media, passing for the quantum mechanics (QM), quantum field theory
(QFT)\ and relativistic astrophysics, the correct description of the physical
phenomena is based in the application of the geometry and group theory.
Between these methods the introduction of spinors constrained by conformal
symmetries in order to map the spacetime, have a particular importance in
classical and quantum field theories. For example, in classical mechanics,
string theories and supersymmetrical ectensions of the spacetime, this
construction (map) has been successfully introduced for the description of
different scenarios. Here we introduce such "spinorial mapping"
(supertwistors) to describe the diverse superspaces. Accordlying with
Penrose's suggestion, the spacetime continuum can be considered as a
derivative construction with respect \ to an underlying spinor structure. For
instance, the spinor structure contains the pre-images of the fundamental
properties of the classical spacetime: dimension, signature, connections, etc.
These superspaces can be the basis of nonlinear realized unified theories
containing the SM+GR (standard model + general relativity) that, with the help
of a super biquaternionic extension of the coordinates, the correct number of
fields will be reached.

\section{Twistor theory and quaternionic extension}

As suggested by Penrose long ago\cite{pen}, from the beginning in the twistor
theory the starting point a complex space $\mathbb{C}M\sim\mathbb{C}%
_{2,4}\left(  T\right)  $ by mean conformal spinors $t_{\alpha}=\left(
\omega^{\overset{\cdot}{\alpha}},\pi_{\alpha}\right)  $ with $\alpha=1,2$ and
$a=1,2,3,4$ as describing the prior geometry with the complex Minkowski space
coordinates usually denoted
\begin{equation}
z^{\overset{\cdot}{\alpha}\beta}=\frac{1}{2}\sigma_{\mu}^{\overset{\cdot
}{\alpha}\beta}z^{\mu} \tag{1}%
\end{equation}
related with the twistors coordinates $t_{\alpha}\subset T$ by the incidence
equation%
\begin{equation}
\omega^{\overset{\cdot}{\alpha}}=iz^{\overset{\cdot}{\alpha}\beta}\pi_{\beta}
\tag{2}%
\end{equation}
that is in fact, a particular case of geometrical (in general harmonic)
mapping. Notice that (1) is directly a biquaternion, namely%
\begin{equation}
z^{\overset{\cdot}{\alpha}\beta}=\frac{1}{2}\left(  \sigma_{0}^{\overset
{\cdot}{\alpha}\beta}z^{0}+\sigma_{1}^{\overset{\cdot}{\alpha}\beta}%
z^{1}+\sigma_{2}^{\overset{\cdot}{\alpha}\beta}z^{2}+\sigma_{3}^{\overset
{\cdot}{\alpha}\beta}z^{3}\right)  \text{ ,\ \ \ \ }z^{\mu}\in\mathbb{C}
\tag{3}%
\end{equation}
that have 8 real dimensions and can be extended even more to 16 real
dimensional if each coordinate is quaternionic itself, namely
\begin{equation}
q^{\overset{\cdot}{\alpha}\beta}=\frac{1}{2}\left(  \sigma_{0}^{\overset
{\cdot}{\alpha}\beta}q^{0}+\sigma_{1}^{\overset{\cdot}{\alpha}\beta}%
q^{1}+\sigma_{2}^{\overset{\cdot}{\alpha}\beta}q^{2}+\sigma_{3}^{\overset
{\cdot}{\alpha}\beta}q^{3}\right)  \text{ ,\ \ \ \ }q^{\mu}\in\mathbb{H}
\tag{4}%
\end{equation}
Now we can promote $t_{\alpha}=\left(  \omega^{\overset{\cdot}{\alpha}}%
,\pi_{\alpha}\right)  $ to fermionic quaternion then a quaternionic twistor is
a quaternion with conformal spinors as coefficients. From (2) a point in a CM
space (in this case an element of $\mathbb{H}_{1}\left(  \mathbb{C}\right)  )$
defines a plane in T or a line in $CP\left(  3\right)  $. Consequently the
mapping (1) is a 1-dimensional quaternionic one (minimal map in $\mathbb{H}$).\ 

\subsection{Quaternionic conformal spinors}

A 4 component spinor over a field $K$ can be realized following the scheme:

\begin{remark}%
\begin{align}
&
\begin{array}
[c]{cccc}%
Majorana & :(K=\mathbb{R)} & ,\xi\left(  x\right)   &
\end{array}
\tag{5}\\
&
\begin{array}
[c]{ccc}%
Dirac & :(K=\mathbb{C)} & ,\psi\left(  x\right)  =\frac{1}{\sqrt{2}}\left(
\xi_{1}\left(  x\right)  +i\xi_{2}\left(  x\right)  \right)
\end{array}
\tag{6}\\
&
\begin{array}
[c]{cccc}%
Quaternionic & :\left(  K=\mathbb{H}\right)   & ,\Psi\left(  x\right)
=\frac{1}{\sqrt{2}}\left(  \xi_{0}\left(  x\right)  +\widehat{i}_{i}\xi
_{i}\left(  x\right)  \right)   & \left(  i=1,2,3\right)
\end{array}
\tag{7}\\
&
\begin{array}
[c]{cccc}%
bi-Quaternionic & :\left(  K=\mathbb{H}_{\mathbb{C}}\right)   & ,\Psi\left(
x\right)  =\frac{1}{\sqrt{2}}\left(  \psi_{0}\left(  x\right)  +\widehat
{i}_{i}\psi_{i}\left(  x\right)  \right)   & \left(  i=1,2,3\right)
\end{array}
\tag{8}%
\end{align}
Case 5 is an ordinary Majorana fermion realized over reals, Case 6 is the
complex realized Dirac one, Case 7 is the quaternionic Dirac field (ordinary
fermionic-quaternion) and the 8 case is a biquaternionic realized spinor where
each coefficient is a Dirac field. The Case 8 is what we are interested in.
\end{remark}

\subsection{Quaternionic extension}

In the quaternionic-twistor theory our starting point a quaternionic space
$\mathbb{H}M\sim\mathbb{H}_{1,4}\left(  T\right)  $ implemented as
$\mathbb{C}M\times\mathbb{C}M$ by mean quaternionic spinors $t_{\alpha
}=\left(  \omega^{\overset{\cdot}{\alpha}},\pi_{\alpha}\right)  $ (with
$\alpha=1,2$ and $a=1,2,3,4$ and $\omega^{\overset{\cdot}{\alpha}},\pi
_{\alpha}\in\mathbb{H)}$as describing the prior geometry with the quaternionic
Minkowski space coordinates denoted
\begin{equation}
q^{\overset{\cdot}{\alpha}\beta}=\frac{1}{2}\sigma_{\mu}^{\overset{\cdot
}{\alpha}\beta}q^{\mu} \tag{9}%
\end{equation}
related with the twistors coordinates $t_{\alpha}\subset T$ by the
quaternionic incidence equation%
\begin{equation}
\omega^{\overset{\cdot}{\alpha}}=iq^{\overset{\cdot}{\alpha}\beta}\pi_{\beta}
\tag{10}%
\end{equation}
we have now 16 real dimensions being each coordinate quaternionic
\begin{equation}
q^{\overset{\cdot}{\alpha}\beta}=\frac{1}{2}\left(  \sigma_{0}^{\overset
{\cdot}{\alpha}\beta}q^{0}+\sigma_{1}^{\overset{\cdot}{\alpha}\beta}%
q^{1}+\sigma_{2}^{\overset{\cdot}{\alpha}\beta}q^{2}+\sigma_{3}^{\overset
{\cdot}{\alpha}\beta}q^{3}\right)  \text{ ,\ \ \ \ }q^{\mu}\in\mathbb{H}
\tag{11}%
\end{equation}
being $\omega^{\overset{\cdot}{\alpha}},\pi_{\alpha}\in\mathbb{H}$
biquaternionic fermions of the form given by the Case 8 (see\cite{luk} for the
simplest case with reality condition), namely $%
\begin{array}
[c]{cc}%
\Psi\left(  x\right)  =\frac{1}{\sqrt{2}}\left(  \psi_{0}\left(  x\right)
+\widehat{i}_{i}\psi_{i}\left(  x\right)  \right)  & \left(  i=1,2,3\right)
\end{array}
,$ as we have pointed out before. As in the ordinary twistor case, we can
introduce a pair of non parallel quaternionic twistors to determine the
corresponding "point" (really a $\mathbb{R}$-subspace) $q\in\mathbb{H}M$ by
solving the matrix equation
\begin{align}
Q  &  =-i\Omega\Pi^{-1}\rightarrow\tag{12}\\
&  \rightarrow q^{\overset{\cdot}{\alpha}\beta}=-i\frac{\left(  \omega
_{1}^{\overset{\cdot}{\alpha}}\pi_{2}^{\beta}-\omega_{2}^{\overset{\cdot
}{\alpha}}\pi_{1}^{\beta}\right)  }{\pi_{2}^{\alpha}\pi_{\alpha1}}\nonumber
\end{align}
where the matrices are $Q=\left(  q^{\overset{\cdot}{\alpha}\beta}\right)
,\Omega=\left(  \omega_{1}^{\overset{\cdot}{\alpha}},\omega_{2}^{\overset
{\cdot}{\alpha}}\right)  ,\Pi=\left(  \pi_{\alpha1,}\pi_{\alpha2}\right)  $

As is easily seen, the $\mathbb{H}M$\ coordinates are invariant under the
transformations in T as follows%
\begin{equation}
\Omega^{\prime}=X\Omega,\text{ \ \ }\Pi^{\prime}=X\text{\ }\Pi\tag{13}%
\end{equation}
with $X\subset GL\left(  2,\mathbb{H}\right)  .$ Notice the important fact
that the introduction of biquaternions (e.g. an almost complex structure)
imply an underlying symplectic one.

\section{Supergroups and quaternionic superframes}

Some points to consider about complex graded vector spaces and quaternionic
ones :

i) for quaternions, the minimal dimension for its representation is 2 (
e.g.$SU\left(  2\right)  \times\mathbb{R}_{1})$

ii) the corresponding complex graded vector space is consequently
$\mathbb{C}^{2n;2m}$ equivalent to $\mathbb{H}^{n;m}$ due i) with n
quaternions with $4n$ complex commuting coefficients (bosons) and $4m$ complex
anticommuting coefficients (fermions)

\subsection{Fundamental representations}

As was commented somewhere for the simplest complex case \cite{luno} (see also
\cite{mori} for introduction to quaternionic structures in QFT), it is
possible to introduce (in the biquaternionic case) the following two
fundamental representations of $SU\left(  2,2;N\left\Vert \mathbb{H}%
_{\mathbb{C}}\right.  \right)  :$

a) Quaternionic supertwistors:%
\begin{align}
T_{A}^{\left(  N\right)  }  &  =(t_{1}....t_{4},\xi_{1}.....\xi_{N})\subset
T_{\mathbb{H}_{\mathbb{C}}}^{\left(  N\right)  }\equiv\mathbb{H}_{\mathbb{C}%
}^{4;N}\tag{14}\\
&  \left(  \mathbb{C}^{8;2N}\text{biquaternionic extension of the Ferber
construction}\right) \nonumber
\end{align}

b) Quaternionic fermionic supertwistors%

\begin{align}
\widetilde{T}_{A}^{\left(  N\right)  }  &  =(\eta_{1}....\eta_{4}%
,u_{1}.....u_{N})\subset\widetilde{T}_{\mathbb{H}_{\mathbb{C}}}^{\left(
N\right)  }\equiv\mathbb{H}_{\mathbb{C}}^{N;4}\tag{15}\\
&  \left(  \mathbb{C}^{2N;8}\text{biquaternionic extension of the
Litov-Pervushin construction}\right) \nonumber
\end{align}
For these two biquaternionic representations the following $U\left(
2,2;N\left\Vert \mathbb{H}_{\mathbb{C}}\right.  \right)  $ scalar products can
be introduced:

$\left\langle T^{\left(  N\right)  },T^{\left(  N\right)  ^{\prime}%
}\right\rangle =\overline{T}_{A}^{\left(  N\right)  }G^{AB}T_{B}^{\left(
N\right)  \prime}(even)$

$\left\langle \widetilde{T}^{\left(  N\right)  },\widetilde{T}^{\left(
N\right)  ^{\prime}}\right\rangle =\overline{\widetilde{T}}_{A}^{\left(
N\right)  }G^{AB}\widetilde{T}_{B}^{\left(  N\right)  \prime}(even)$

$\left\langle \widetilde{T}^{\left(  N\right)  },T^{\left(  N\right)
^{\prime}}\right\rangle =\left\langle T^{\left(  N\right)  },\widetilde
{T}^{\left(  N\right)  ^{\prime}}\right\rangle =\overline{\widetilde{T}}%
_{A}^{\left(  N\right)  }G^{AB}T_{B}^{\left(  N\right)  \prime}=\overline
{T}_{A}^{\left(  N\right)  }G^{AB}\widetilde{T}_{B}^{\left(  N\right)  \prime
}(odd:$ linear in fermionic coordinates)

where $G_{AB}^{\left(  N\right)  }$ is the $U\left(  2,2;N\left\Vert
\mathbb{H}_{\mathbb{C}}\right.  \right)  $ supermetric, schematically
$G_{AB}^{\left(  N\right)  }=\left(
\begin{array}
[c]{cc}%
g & 0\\
0 & iI_{N}%
\end{array}
\right)  \ g$ is symplectic and each entry is a quaternionic one (e.g a
$2\times2$ block). Consequently the supergroup $U\left(  2,2;N\left\Vert
\mathbb{H}_{\mathbb{C}}\right.  \right)  $ is defined as the set of graded
$2\left(  4+N\right)  \times2\left(  4+N\right)  $ matrices $U$ ($\left(
4+N\right)  \times\left(  4+N\right)  $ biquaternionic matrices in the lowest
$\mathbb{H}$-representation) satisfiying the relation
\[
^{\ast}U_{DA}G_{AB}^{\left(  N\right)  }U_{BC}=G_{DC}^{\left(  N\right)  }%
\]
Let us to observe, in resume, the following points:

\begin{itemize}
\item A) the graded quaternionic matrices $U$ :
\end{itemize}

i) are described by four \textit{biquaternionic supertwistors} and $N$
\textit{biquaternionic fermionic supertwistors};

ii) are represented with $2\left(  4+N\right)  \times2\left(  4+N\right)  $
matrices $U$ ($\left(  4+N\right)  \times\left(  4+N\right)  $ biquaternionic
matrices in the lowest $\mathbb{H}$-representation)

\begin{itemize}
\item B) and the four \textit{biquaternionic supertwistors} and $N$
\textit{biquaternionic fermionic supertwistors} satisfy:
\end{itemize}

i) $16=4\times4$ relations defining an arbitrary four frame in the
biquaternionic supertwistor space $\mathbb{H}_{\mathbb{C}}^{4;N};$

ii) $N\times N$ \ biquaternionic relations defining an arbitrary $N$-frame in
the biquaternionic fermionic supertwistor space $\mathbb{H}_{\mathbb{C}}%
^{N;N};$

iii) $4N$ biquaternionic relations defining graded structure (orthogonality in
some cases) of the biquaternionic superframes in $\mathbb{H}_{\mathbb{C}%
}^{4;N}$ and $\mathbb{H}_{\mathbb{C}}^{N;4}.$Schematically these structures
are:%
\[
\left(
\begin{array}
[c]{cc}%
4\times4 & 4\times N\left(  \mathbb{H}_{\mathbb{C}}^{4,N}\right) \\
N\times4\left(  \mathbb{H}_{\mathbb{C}}^{N,4}\right)  & N\times N\left(
\mathbb{H}_{\mathbb{C}}^{N,N}\right)
\end{array}
\right)
\]

\begin{itemize}
\item C) and the most important is that the biquaternionic fields $t,\xi
,\eta,u$ (indexes avoided) into the structures $T_{A}^{\left(  N\right)  }$
and $\widetilde{T}_{A}^{\left(  N\right)  }$contain four spinors each one as
coefficients: e.g.
\begin{align}
t_{a}  &  =e_{0}t_{a}^{0}+e_{i}t_{a}^{i},\tag{16}\\
u_{N}  &  =e_{0}u_{N}^{0}+e_{i}u_{N}^{i},(i=1,2,3) \tag{17}%
\end{align}
etc. In consequence, the supergroup $SU\left(  2,2;N\mid\mathbb{H}%
_{\mathbb{C}}\right)  $ \ describes all the superframes-( modulo a global
phase factor) in $\mathcal{T}_{\mathbb{H}_{\mathbb{C}}}^{\left(  N\right)
}=T_{\mathbb{H}_{\mathbb{C}}}^{\left(  N\right)  }\oplus\widetilde
{T}_{\mathbb{H}_{\mathbb{C}}}^{\left(  N\right)  }=\mathbb{H}_{\mathbb{C}%
}^{4;N}\oplus\mathbb{H}_{\mathbb{C}}^{N;4}$(Penrose notation), being the
corresponding parametric equations describing the respective coset written in
$\mathcal{T}_{\mathbb{H}_{\mathbb{C}}}^{\left(  N\right)  }.$
\end{itemize}

\section{Penrose equations and super-quaternionic extension: the general case}

The quaternionic superspace coordinates are $\left(  N=p+k\right)  $%
\begin{equation}
\mathbb{H}_{\mathbb{C}}\left(  N,N-k\right)  =\left(  q_{\mu},\lambda
_{s}^{\text{ \ }r},\theta_{s}^{\text{ \ }\alpha},\theta^{\overset{\cdot
}{\alpha}r}\right)  \tag{18}%
\end{equation}
where in the general case, $k(N-k)$ bosonic quaternionic coordinates describe
the internal symmetry coset $\frac{U\left(  N\mid\mathbb{H}_{\mathbb{C}%
}\right)  }{U\left(  N\mid\mathbb{H}_{\mathbb{C}}\right)  U\left(
N-k\mid\mathbb{H}_{\mathbb{C}}\right)  }$ and shall be called \textit{regular}
or \textit{Fueter-analytic} coordinates. \textbf{The above quaternionic
superspace coordinate can be determined in terms of two supertwistor and k
fermionic supertwistors}. This can be achieved introducing the following
graded matrices:

i) the $\left(  2+N-k\right)  \times\left(  2+k\right)  $ quaternionic matrix
describing the quaternionic superspace coordinates%
\begin{equation}
Q=\left(
\begin{array}
[c]{cc}%
q^{\overset{\cdot}{\alpha}\beta} & \theta^{\overset{\cdot}{\alpha}r}\\
\theta_{s}^{\text{ \ }\beta} & \lambda_{s}^{\text{ \ }r}%
\end{array}
\right)  \tag{19}%
\end{equation}
also the graded $\left(  2+N-k\right)  $ $\times\left(  2+k\right)  $
quaternionic matrix
\begin{equation}
\Omega=\left(
\begin{array}
[c]{cc}%
\omega_{\text{ }\gamma}^{\overset{\cdot}{\alpha}} & \rho_{\text{ }r}%
^{\overset{\cdot}{\alpha}}\\
\varsigma_{s\gamma} & u_{sr}%
\end{array}
\right)  \tag{20}%
\end{equation}
and the $\left(  2+k\right)  \times\left(  2+k\right)  $ quaternionic matrix
\begin{equation}
\Pi=\left(
\begin{array}
[c]{cc}%
\pi_{\alpha\gamma} & \sigma_{\alpha r^{\prime}}\\
\xi_{r\gamma} & v_{rr^{\prime}}%
\end{array}
\right)  \tag{21}%
\end{equation}
where $\gamma=1,2$ is twistor index and $r=1,....k$, $s=1,....N-k$
$\ \ \ \left(  N-k=p\text{ in the Litov-Pervushin notation}\right)
.$Consequently, the superextension of the ordinary twistor equations becomes
to%
\begin{equation}
\left(
\begin{array}
[c]{cc}%
\omega_{\text{ }\gamma}^{\overset{\cdot}{\alpha}} & \rho_{\text{ }r^{\prime}%
}^{\overset{\cdot}{\alpha}}\\
\varsigma_{s\gamma} & u_{sr^{\prime}}%
\end{array}
\right)  =\left(
\begin{array}
[c]{cc}%
q^{\overset{\cdot}{\alpha}\beta}\pi_{\beta\gamma}+\theta^{\overset{\cdot
}{\alpha}r}\xi_{r\gamma} & q^{\overset{\cdot}{\alpha}\beta}\sigma_{\beta
r^{\prime}}+\theta^{\overset{\cdot}{\alpha}r}v_{rr^{\prime}}\\
\theta_{s}^{\text{ \ }\beta}\pi_{\beta\gamma}+\lambda_{s}^{\text{ \ }r}%
\xi_{r\gamma} & \theta_{s}^{\text{ \ }\beta}\sigma_{\beta r^{\prime}}%
+\lambda_{s}^{\text{ \ }r}v_{rr^{\prime}}%
\end{array}
\right)  \tag{22}%
\end{equation}
Consequently, the reconstruction via twistors of the superspace namely
\begin{equation}
Q=-i\Omega\Pi^{-1} \tag{23}%
\end{equation}
using%

\begin{equation}
\Pi^{-1}=\left(
\begin{array}
[c]{cc}%
A^{\gamma\beta} & B^{\gamma^{\prime}r}\\
C^{\gamma\beta} & D^{r^{\prime}r}%
\end{array}
\right)  \tag{24}%
\end{equation}
where
\begin{equation}%
\begin{array}
[c]{ll}%
A^{\gamma\beta}=-\left(  \pi^{-1}\right)  ^{r^{\prime\prime}\gamma}%
\sigma_{r^{\prime\prime}r^{\prime}}\left[  v^{r^{\prime}\beta}-\xi_{\text{
}\gamma^{\prime}}^{\beta}\left(  \pi^{-1}\right)  ^{\gamma^{\prime\prime
}\gamma^{\prime}}\sigma_{\gamma^{\prime\prime}}^{\text{ \ }r^{\prime}}\right]
^{-1} & ,B^{\gamma^{\prime}r}=-\left(  \xi^{-1}\right)  ^{r^{\prime\prime
}\gamma^{\prime}}v_{r^{\prime\prime}r^{\prime}}\left[  \sigma^{rr^{\prime}%
}-\pi^{r\gamma}\left(  \xi^{-1}\right)  _{\gamma r^{\prime\prime\prime}%
}v^{r^{\prime\prime\prime}r^{\prime}}\right]  ^{-1}\\
C^{r^{\prime}\beta}=\left[  v^{r^{\prime}\beta}-\xi_{\text{ }\gamma^{\prime}%
}^{\beta}\left(  \pi^{-1}\right)  ^{\gamma^{\prime\prime}\gamma^{\prime}%
}\sigma_{\gamma^{\prime\prime}}^{\text{ \ }r^{\prime}}\right]  ^{-1}, &
D^{r^{\prime}r}=\left[  \sigma^{rr^{\prime}}-\pi_{\text{ }\gamma}^{r}\left(
\xi^{-1}\right)  ^{\gamma r^{\prime\prime\prime}}v_{r^{\prime\prime\prime}%
}^{\text{ }r^{\prime}}\right]  ^{-1}%
\end{array}
\tag{25}%
\end{equation}
carry explicitly to the following expressions in general form:%
\begin{align}
q^{\overset{\cdot}{\alpha}\beta}  &  =\left(  -\omega_{\text{ }\gamma
}^{\overset{\cdot}{\alpha}}\left(  \pi^{-1}\right)  ^{\gamma^{\prime}\gamma
}\sigma_{\gamma^{\prime}r^{\prime}}+\rho_{\text{ }r^{\prime}}^{\overset{\cdot
}{\alpha}}\right)  \left[  v^{r^{\prime}\beta}-\xi_{\text{ }\gamma^{\prime}%
}^{\beta}\left(  \pi^{-1}\right)  ^{\gamma^{\prime\prime}\gamma^{\prime}%
}\sigma_{\gamma^{\prime\prime}}^{\text{ \ }r^{\prime}}\right]  ^{-1}%
,\tag{26}\\
\theta^{\overset{\cdot}{\alpha}r}  &  =\left(  -\omega_{\text{ }\gamma
}^{\overset{\cdot}{\alpha}}\left(  \xi^{-1}\right)  ^{\gamma^{\prime}\gamma
}v_{\gamma^{\prime}r^{\prime}}+\rho_{\text{ }r^{\prime}}^{\overset{\cdot
}{\alpha}}\right)  \left[  \sigma^{rr^{\prime}}-\pi_{\text{ }\gamma}%
^{r}\left(  \xi^{-1}\right)  ^{\gamma r^{\prime\prime\prime}}v_{r^{\prime
\prime\prime}}^{\text{ }r^{\prime}}\right]  ^{-1}\tag{27}\\
\theta_{s}^{\text{ \ }\beta}  &  =\left(  -\varsigma_{s\gamma}\left(  \pi
^{-1}\right)  ^{\gamma^{\prime}\gamma}\sigma_{\gamma^{\prime}r^{\prime}%
}+u_{sr^{\prime}}\right)  \left[  v^{r^{\prime}\beta}-\xi_{\text{ }%
\gamma^{\prime}}^{\beta}\left(  \pi^{-1}\right)  ^{\gamma^{\prime\prime}%
\gamma^{\prime}}\sigma_{\gamma^{\prime\prime}}^{\text{ \ }r^{\prime}}\right]
^{-1}\tag{28}\\
\lambda_{s}^{\text{ \ }r}  &  =\left(  -\varsigma_{s\gamma}\left(  \xi
^{-1}\right)  ^{\gamma^{\prime}\gamma}v_{\gamma^{\prime}r^{\prime}%
}+u_{sr^{\prime}}\right)  \left[  \sigma^{rr^{\prime}}-\pi_{\text{ }\gamma
}^{r}\left(  \xi^{-1}\right)  ^{\gamma r^{\prime\prime\prime}}v_{r^{\prime
\prime\prime}}^{\text{ }r^{\prime}}\right]  ^{-1} \tag{29}%
\end{align}

\section{The N=4, k=2 superspace}

This is the corresponding to the case $p=q=2$ $(N=p+q)$ in the Litov
-Pervushin notation\cite{Litov-84}. Eqs.(23) and (24) describe for N=4, k=2
the supercoset%
\begin{equation}
\frac{SU\left(  2,2;4\mid\mathbb{H}_{\mathbb{C}}\right)  }{SU\left(
2;\underset{k}{\underbrace{2}}\mid\mathbb{H}_{\mathbb{C}}\right)  \times
SU\left(  2;\underset{N-k=p}{\underbrace{2}}\mid\mathbb{H}_{\mathbb{C}%
}\right)  } \tag{30}%
\end{equation}%
\begin{equation}
Q=\left(
\begin{array}
[c]{cccc}%
q^{\overset{\cdot}{1}1} & q^{\overset{\cdot}{1}2} & \theta^{\overset{\cdot}%
{1}1} & \theta^{\overset{\cdot}{1}2}\\
q^{\overset{\cdot}{2}1} & q^{\overset{\cdot}{2}2} & \theta^{\overset{\cdot}%
{2}1} & \theta^{\overset{\cdot}{2}2}\\
\theta_{1}^{\text{ \ }1} & \theta_{1}^{\text{ \ }2} & \lambda_{1}^{\text{
\ }1} & \lambda_{1}^{\text{ \ }2}\\
\theta_{2}^{\text{ \ }1} & \theta_{2}^{\text{ \ }2} & \lambda_{2}^{\text{
\ }1} & \lambda_{2}^{\text{ \ }2}%
\end{array}
\right)  \tag{31}%
\end{equation}%
\begin{equation}
\Pi=\left(
\begin{array}
[c]{cccc}%
\pi_{11} & \pi_{12} & \sigma_{11} & \sigma_{12}\\
\pi_{21} & \pi_{22} & \sigma_{21} & \sigma_{22}\\
\xi_{11} & \xi_{12} & v_{11} & v_{12}\\
\xi_{21} & \xi_{22} & v_{21} & v_{22}%
\end{array}
\right)  \tag{32}%
\end{equation}%
\begin{equation}
\Omega=\left(
\begin{array}
[c]{cccc}%
\omega_{1}^{\overset{\cdot}{1}} & \omega_{2}^{\overset{\cdot}{1}} & \rho
_{1}^{\overset{\cdot}{1}} & \rho_{2}^{\overset{\cdot}{1}}\\
\omega_{1}^{\overset{\cdot}{2}} & \omega_{2}^{\overset{\cdot}{2}} & \rho
_{1}^{\overset{\cdot}{2}} & \rho_{2}^{\overset{\cdot}{2}}\\
\varsigma_{11} & \varsigma_{12} & u_{11} & u_{12}\\
\varsigma_{12} & \varsigma_{22} & u_{12} & u_{22}%
\end{array}
\right)  \tag{33}%
\end{equation}
where $\gamma=1,2$ is twistor index $r=1,....k$, $s=1,....N-k$ $\ \ \ \left(
N-k=p\right)  $

Explicitly%
\begin{align}
q^{\overset{\cdot}{\alpha}\beta}  &  =\left(  -\omega_{\text{ }1}%
^{\overset{\cdot}{\alpha}}\left(  \pi^{-1}\right)  ^{21}\sigma_{2r^{\prime}%
}+\omega_{\text{ }2}^{\overset{\cdot}{\alpha}}\left(  \pi^{-1}\right)
^{12}\sigma_{1r^{\prime}}+\rho_{\text{ }r^{\prime}}^{\overset{\cdot}{\alpha}%
}\right)  \left[  v^{r^{\prime}\beta}-\xi_{\text{ }1}^{\beta}\left(  \pi
^{-1}\right)  ^{21}\sigma_{2}^{\text{ \ }r^{\prime}}+\xi_{\text{ }2}^{\beta
}\left(  \pi^{-1}\right)  ^{12}\sigma_{1}^{\text{ \ }r^{\prime}}\right]
^{-1},\tag{34}\\
\theta^{\overset{\cdot}{\alpha}r}  &  =\left(  -\omega_{\text{ }1}%
^{\overset{\cdot}{\alpha}}\left(  \xi^{-1}\right)  ^{21}v_{2r^{\prime}}%
+\omega_{\text{ }2}^{\overset{\cdot}{\alpha}}\left(  \xi^{-1}\right)
^{12}v_{1r^{\prime}}+\rho_{\text{ }r^{\prime}}^{\overset{\cdot}{\alpha}%
}\right)  \left[  \sigma^{rr^{\prime}}-\pi_{\text{ }1}^{r}\left(  \xi
^{-1}\right)  _{2}^{\text{ \ }r^{\prime\prime\prime}}v_{r^{\prime\prime\prime
}}^{\text{ }r^{\prime}}+\pi_{\text{ }2}^{r}\left(  \xi^{-1}\right)
_{1}^{\text{ \ }r^{\prime\prime\prime}}v_{r^{\prime\prime\prime}}^{\text{
}r^{\prime}}\right]  ^{-1}\tag{35}\\
\theta_{s}^{\text{ \ }\beta}  &  =\left(  -\varsigma_{s1}\left(  \pi
^{-1}\right)  ^{21}\sigma_{2r^{\prime}}+\varsigma_{s2}\left(  \pi^{-1}\right)
^{12}\sigma_{1r^{\prime}}+u_{sr^{\prime}}\right)  \left[  v^{r^{\prime}\beta
}-\xi_{\text{ }1}^{\beta}\left(  \pi^{-1}\right)  ^{21}\sigma_{2}^{\text{
\ }r^{\prime}}+\xi_{\text{ }2}^{\beta}\left(  \pi^{-1}\right)  ^{12}\sigma
_{1}^{\text{ \ }r^{\prime}}\right]  ^{-1}\tag{36}\\
\lambda_{s}^{\text{ \ }r}  &  =\left(  -\varsigma_{s1}\left(  \xi^{-1}\right)
^{21}v_{2r^{\prime}}+\varsigma_{s2}\left(  \xi^{-1}\right)  ^{12}%
v_{1r^{\prime}}+u_{sr^{\prime}}\right)  \left[  \sigma^{rr^{\prime}}%
-\pi_{\text{ }1}^{r}\left(  \xi^{-1}\right)  _{2}^{\text{ \ }r^{\prime
\prime\prime}}v_{r^{\prime\prime\prime}}^{\text{ }r^{\prime}}+\pi_{\text{ }%
2}^{r}\left(  \xi^{-1}\right)  _{1}^{\text{ \ }r^{\prime\prime\prime}%
}v_{r^{\prime\prime\prime}}^{\text{ }r^{\prime}}\right]  ^{-1} \tag{37}%
\end{align}
Evidently, the above solutions are invariant under the arbitrary $R\in
GL\left(  2;2\mid\mathbb{H}_{\mathbb{C}}\right)  $ supertransformations (e.g.:
superrotations) namely%
\begin{equation}
\Omega^{\prime}\rightarrow\Omega R,\text{ \ \ \ }\Pi^{\prime}\rightarrow\Pi
R\text{\ } \tag{38}%
\end{equation}
with $\Omega$ and $\Pi$ for $N=4,k=2.$ \textbf{We easily see that two
biquaternionic supertwistors and two biquaternionic fermionic supertwistors
are described by sixty four bosonic complex coordinates (given by eigth
bosonic biquaternions )and by sixty four fermionic complex coordinates (given
by eigth fermionic biquaternions): one half of these coordinates describes the
}$\mathbf{N=4,k=2}$\textbf{ biquaternionic superspace and the second half,
however, describes the }$GL\left(  2;2\mid\mathbb{H}_{\mathbb{C}}\right)  $
degrees of freedom. Schematically, from the matrix $Q$, each block describes
faithfullly the following supercoordinates%
\begin{align}
\left(
\begin{array}
[c]{cc}%
q^{\overset{\cdot}{1}1} & q^{\overset{\cdot}{1}2}\\
q^{\overset{\cdot}{2}1} & q^{\overset{\cdot}{2}2}%
\end{array}
\right)   &  \rightarrow2\times4\times4=32\text{ bosonic fields}\tag{39}\\
\left(
\begin{array}
[c]{cc}%
\lambda_{1}^{\text{ \ }1} & \lambda_{1}^{\text{ \ }2}\\
\lambda_{2}^{\text{ \ }1} & \lambda_{2}^{\text{ \ }2}%
\end{array}
\right)   &  \rightarrow2\times4\times4=32\text{ bosonic fields} \tag{40}%
\end{align}%
\begin{align}
\left(
\begin{array}
[c]{cc}%
\theta_{1}^{\text{ \ }1} & \theta_{1}^{\text{ \ }2}\\
\theta_{2}^{\text{ \ }1} & \theta_{2}^{\text{ \ }2}%
\end{array}
\right)   &  \rightarrow2\times4\times4=32\text{ fermionic fields}\tag{41}\\
\left(
\begin{array}
[c]{cc}%
\theta^{\overset{\cdot}{1}1} & \theta^{\overset{\cdot}{1}2}\\
\theta^{\overset{\cdot}{2}1} & \theta^{\overset{\cdot}{2}2}%
\end{array}
\right)   &  \rightarrow2\times4\times4=32\text{ fermionic fields} \tag{42}%
\end{align}

It is important to remark here that if the supermanifold described (spanned)
by two biquaternionic supertwistors : $T_{1}^{\left(  4\right)  }$
$T_{2}^{\left(  4\right)  }$and two biquaternionic fermionic$\widetilde{T}%
_{1}^{\left(  4\right)  }$ $\widetilde{T}_{2}^{\left(  4\right)  }%
$supertwistors is totally null or supergeodesic with respect to the norm of
$SU\left(  2,2;4,4\left\Vert \mathbb{H}_{\mathbb{C}}\right.  \right)  ,$
namely%
\begin{gather}
\left\langle T_{1}^{\left(  4\right)  },T_{1}^{\left(  4\right)
}\right\rangle =\left\langle T_{2}^{\left(  4\right)  },T_{2}^{\left(
4\right)  }\right\rangle =\left\langle \widetilde{T}_{1}^{\left(  4\right)
},\widetilde{T}_{1}^{\left(  4\right)  }\right\rangle =\left\langle
\widetilde{T}_{2}^{\left(  4\right)  },\widetilde{T}_{2}^{\left(  4\right)
}\right\rangle =0\tag{43}\\
\left\langle T_{1}^{\left(  4\right)  },T_{2}^{\left(  4\right)
}\right\rangle =\left\langle T_{2}^{\left(  4\right)  },\widetilde{T}%
_{1}^{\left(  4\right)  }\right\rangle =\left\langle T_{1}^{\left(  4\right)
},\widetilde{T}_{2}^{\left(  4\right)  }\right\rangle =\tag{44}\\
=\left\langle \widetilde{T}_{1}^{\left(  4\right)  },T_{1}^{\left(  4\right)
}\right\rangle =\left\langle T_{2}^{\left(  4\right)  },\widetilde{T}%
_{2}^{\left(  4\right)  }\right\rangle =\left\langle \widetilde{T}%
_{1}^{\left(  4\right)  },\widetilde{T}_{2}^{\left(  4\right)  }\right\rangle
=0\nonumber
\end{gather}
notice that this fact doesn't implies, in principle, a reality condition.
However, these conditions enforce a Majorana condition over the antidiagonal
fermionic sectors into the matrix $Q.$

Notice that the constraints (54) four bosonic complex variables can be
eliminated, and the constraints (55) two complex bosonic variables and four
fermionic ones can be eliminated. This counting evisently agrees with the
restriction of biquaternionic superspace degrees of freedom of by the graded
pseudohermiticity condition given by construction.

\section{The N=8, k=4 superspace}

This is the corresponding to the case $p=q=2$ $(N=p+q)$ in the Litov
-Pervushin notation. Eqs.(23) and (24) describe for $N=8$, $k=4$ the
supercoset%
\begin{equation}
\frac{SU\left(  2,2;8\mid\mathbb{H}_{\mathbb{C}}\right)  }{SU\left(
2;\underset{k}{\underbrace{4}}\mid\mathbb{H}_{\mathbb{C}}\right)  \times
SU\left(  2;\underset{N-k=p}{\underbrace{4}}\mid\mathbb{H}_{\mathbb{C}%
}\right)  } \tag{45}%
\end{equation}%
\begin{equation}
Q=\left(
\begin{array}
[c]{cccccc}%
q^{\overset{\cdot}{1}1} & q^{\overset{\cdot}{1}2} & \theta^{\overset{\cdot}%
{1}1} & \theta^{\overset{\cdot}{1}2} & \theta^{\overset{\cdot}{1}3} &
\theta^{\overset{\cdot}{1}4}\\
q^{\overset{\cdot}{2}1} & q^{\overset{\cdot}{2}2} & \theta^{\overset{\cdot}%
{2}1} & \theta^{\overset{\cdot}{2}2} & \theta^{\overset{\cdot}{2}3} &
\theta^{\overset{\cdot}{2}4}\\
\theta_{1}^{\text{ \ }1} & \theta_{1}^{\text{ \ }2} & \lambda_{1}^{\text{
\ }1} & \lambda_{1}^{\text{ \ }2} & \lambda_{1}^{\text{ \ }3} & \lambda
_{1}^{\text{ \ }4}\\
\theta_{2}^{\text{ \ }1} & \theta_{2}^{\text{ \ }2} & \lambda_{2}^{\text{
\ }1} & \lambda_{2}^{\text{ \ }2} & \lambda_{2}^{\text{ \ }3} & \lambda
_{2}^{\text{ \ }4}\\
\theta_{3}^{\text{ \ }1} & \theta_{3}^{\text{ \ }2} & \lambda_{3}^{\text{
\ }1} & \lambda_{3}^{\text{ \ }2} & \lambda_{3}^{\text{ \ }3} & \lambda
_{3}^{\text{ \ }4}\\
\theta_{4}^{\text{ \ }1} & \theta_{4}^{\text{ \ }2} & \lambda_{4}^{\text{
\ }1} & \lambda_{4}^{\text{ \ }2} & \lambda_{4}^{\text{ \ }3} & \lambda
_{4}^{\text{ \ }4}%
\end{array}
\right)  \tag{46}%
\end{equation}%
\begin{equation}
\Pi=\left(
\begin{array}
[c]{cccccc}%
\pi_{11} & \pi_{12} & \sigma_{11} & \sigma_{12} & \sigma_{13} & \sigma_{14}\\
\pi_{21} & \pi_{22} & \sigma_{21} & \sigma_{22} & \sigma_{23} & \sigma_{24}\\
\xi_{11} & \xi_{12} & v_{11} & v_{12} & v_{13} & v_{14}\\
\xi_{21} & \xi_{22} & v_{21} & v_{22} & v_{23} & v_{24}\\
\xi_{31} & \xi_{32} & v_{31} & v_{32} & v_{33} & v_{34}\\
\xi_{41} & \xi_{42} & v_{41} & v_{42} & v_{43} & v_{44}%
\end{array}
\right)  \tag{47}%
\end{equation}%
\begin{equation}
\Omega=\left(
\begin{array}
[c]{cccccc}%
\omega_{1}^{\overset{\cdot}{1}} & \omega_{2}^{\overset{\cdot}{1}} & \rho
_{1}^{\overset{\cdot}{1}} & \rho_{2}^{\overset{\cdot}{1}} & \rho_{3}%
^{\overset{\cdot}{1}} & \rho_{4}^{\overset{\cdot}{1}}\\
\omega_{1}^{\overset{\cdot}{2}} & \omega_{2}^{\overset{\cdot}{2}} & \rho
_{1}^{\overset{\cdot}{2}} & \rho_{2}^{\overset{\cdot}{2}} & \rho_{3}%
^{\overset{\cdot}{2}} & \rho_{4}^{\overset{\cdot}{2}}\\
\varsigma_{11} & \varsigma_{12} & u_{11} & u_{12} & u_{13} & u_{14}\\
\varsigma_{21} & \varsigma_{22} & u_{21} & u_{22} & u_{23} & u_{24}\\
\varsigma_{31} & \varsigma_{32} & u_{31} & u_{32} & u_{33} & u_{34}\\
\varsigma_{41} & \varsigma_{42} & u_{41} & u_{42} & u_{43} & u_{44}%
\end{array}
\right)  \tag{48}%
\end{equation}
where $\gamma=1,2$ is twistor index $r=1,....k$, $s=1,....N-k$ $\ \ \ \left(
N-k=p\right)  .$ \textbf{This case is very important to have the 64*
dimensional sector to reproduce SU(3) }

Evidently, the above solutions are invariant under the arbitrary $R\in
GL\left(  2;2\mid\mathbb{H}_{\mathbb{C}}\right)  $ supertransformations (e.g.:
superrotations) namely%
\begin{equation}
\Omega^{\prime}\rightarrow\Omega R,\text{ \ \ \ }\Pi^{\prime}\rightarrow\Pi
R\text{\ } \tag{49}%
\end{equation}
with $\Omega$ and $\Pi$ given by (47,48)\ with $N=8,k=4.$ \textbf{We easily
see that two biquaternionic supertwistors and two biquaternionic fermionic
supertwistors are described by sixty four bosonic complex coordinates (given
by eigth bosonic biquaternions )and by sixty four fermionic complex
coordinates (given by eigth fermionic biquaternions): one half of these
coordinates describes the }$\mathbf{N=8,k=4}$\textbf{ biquaternionic
superspace and the second half, however, describes the }$GL\left(
2;2\mid\mathbb{H}_{\mathbb{C}}\right)  $ degrees of freedom. Schematically,
from the matrix (46), each block describes faithfullly the following
supercoordinates%
\begin{align}
\left(
\begin{array}
[c]{cc}%
q^{\overset{\cdot}{1}1} & q^{\overset{\cdot}{1}2}\\
q^{\overset{\cdot}{2}1} & q^{\overset{\cdot}{2}2}%
\end{array}
\right)   &  \rightarrow2\times4\times4=32\text{ bosonic sector}\tag{50}\\
\left(
\begin{array}
[c]{cccc}%
\lambda_{1}^{\text{ \ }1} & \lambda_{1}^{\text{ \ }2} & \lambda_{1}^{\text{
\ }3} & \lambda_{1}^{\text{ \ }4}\\
\lambda_{2}^{\text{ \ }1} & \lambda_{2}^{\text{ \ }2} & \lambda_{2}^{\text{
\ }3} & \lambda_{2}^{\text{ \ }4}\\
\lambda_{3}^{\text{ \ }1} & \lambda_{3}^{\text{ \ }2} & \lambda_{3}^{\text{
\ }3} & \lambda_{3}^{\text{ \ }4}\\
\lambda_{4}^{\text{ \ }1} & \lambda_{4}^{\text{ \ }2} & \lambda_{4}^{\text{
\ }3} & \lambda_{4}^{\text{ \ }4}%
\end{array}
\right)   &  \rightarrow2\times4\times16=128\text{ bosonic sector(internal
"soul")} \tag{51}%
\end{align}%
\begin{equation}
\left(
\begin{array}
[c]{cccc}%
\theta^{\overset{\cdot}{1}1} & \theta^{\overset{\cdot}{1}2} & \theta
^{\overset{\cdot}{1}3} & \theta^{\overset{\cdot}{1}4}\\
\theta^{\overset{\cdot}{2}1} & \theta^{\overset{\cdot}{2}2} & \theta
^{\overset{\cdot}{2}3} & \theta^{\overset{\cdot}{2}4}%
\end{array}
\right)  \rightarrow2\times4\times8=64\text{ fermionic sector} \tag{52}%
\end{equation}%
\begin{equation}
\left(
\begin{array}
[c]{cc}%
\theta_{1}^{\text{ \ }1} & \theta_{1}^{\text{ \ }2}\\
\theta_{2}^{\text{ \ }1} & \theta_{2}^{\text{ \ }2}\\
\theta_{3}^{\text{ \ }1} & \theta_{3}^{\text{ \ }2}\\
\theta_{4}^{\text{ \ }1} & \theta_{4}^{\text{ \ }2}%
\end{array}
\right)  \rightarrow2\times4\times8=64\text{ fermionic sector} \tag{53}%
\end{equation}

\section{Discussion}

As is well known, the knowledgement of the Casimir operators of any group is
important: they are needed for the classification of the irreducible
representations of the (extended supersymmetry) algebra. From the
supersymmetic viewpoint, they also can be used to find covariant equations of
motion for superfields\cite{OgyeSok}. If there are not central charges, the
maximal possible internal symmetry group is $U\left(  N\right)  :$in this case
a complete set of Casimirs operators are well know: $P^{2},$ the square of the
superspin vector and the corresponding super-Casimir extensions for U$\left(
N\right)  .$

In the general cases, was claimed that the central charges are needed being
the maximal possible internal group for any $N$ \textit{even }$USp\left(
N\right)  =Sp\left(  N,\mathbb{H}_{\mathbb{C}}\right)  \cap U\left(
N,\mathbb{H}_{\mathbb{C}}\right)  $, however as we have shown, that if $N=p+k$
with $p,k\neq0$ these central charges are nothing more that the genereators
associated to the tensorial representation (not chiral or antichiral) given by
the $\lambda_{s}^{\text{ \ }r}$ parameters into the superspace matrix $Q.$

\section{Superfields and coherent states}

We know that the equation relating the $B_{0}$ and $B_{1}$ parts of the
superspace in the case of supertwistors can be in terms of quaternions into
the form

i) The fundamental representation can be descomposed, in principle, as in the
case of \cite{Litov-84}\ as
\[
\mathcal{U}=t\cdot h
\]

where h is an element of the maximal compact subgroup $S(U(2m)\times U(2n))$
and $t$ of the corresponding coset space $\frac{SU(2,2;k,N-k)}{S(U(2,k)\times
U(2,N-k))}$. Explicitly (see ref.[5])%
\begin{equation}
h=\exp\left[  i\left(
\begin{array}
[c]{cc}%
\chi & 0\\
0 & \varepsilon
\end{array}
\right)  \right]  =\left(
\begin{array}
[c]{cc}%
\mu & 0\\
0 & \upsilon
\end{array}
\right)  \tag{56}%
\end{equation}
and
\begin{equation}
t=\exp\left[  i\left(
\begin{array}
[c]{cc}%
0 & \nu\\
\overline{\nu} & 0
\end{array}
\right)  \right]  =\frac{1}{\sqrt{1-QQ^{\dagger}}}\left(
\begin{array}
[c]{cc}%
\mathbb{I} & Q\\
Q^{\dagger} & \mathbb{I}%
\end{array}
\right)  \tag{57}%
\end{equation}
where the parametrization is given by
\begin{equation}
Q_{\text{ }B}^{A}=\left[  \frac{\tanh\sqrt{\overline{\nu}\nu}}{\sqrt
{\overline{\nu}\nu}}\nu\right]  _{\text{ }B}^{A} \tag{58}%
\end{equation}
because the quaternionic supervariable transforms nonlinearly under
$SU(2,2;k,N-k)$%
\begin{equation}
Q\rightarrow\frac{\alpha Q+\beta}{\gamma Q+\delta}\text{ ;\ \ \ \ \ \ \ }%
\alpha,\beta,\gamma,\delta\in\mathbb{H}_{\mathbb{C}} \tag{59}%
\end{equation}

the superdisplacement operator is precisely
\begin{equation}
D=e^{\eta^{\dagger}Q^{\dagger}\xi^{\dagger}}e^{\left[  \eta^{\dagger}%
\ln\left(  1-QQ^{\dagger}\right)  ^{1/2}\eta-\xi\ln\left(  1-QQ^{\dagger
}\right)  ^{1/2}\xi^{\dagger}\right]  }e^{-\xi Q\eta} \tag{60}%
\end{equation}
where $\eta$ and $\xi$ are the quaternionic oscillator-like supertwistors of
\cite{Litov-84}(e.g.;super quaternionic analog of the standard $a$ and
$a^{\dagger}$ operators), namely
\begin{align}
\xi^{A}  &  =\left(
\begin{array}
[c]{cc}%
a^{c}, & -\xi^{i}%
\end{array}
\right)  ;\text{ }\overline{\xi}_{A}=\left.  \left(
\begin{array}
[c]{c}%
a_{c}^{\dagger}\\
\xi_{i}^{\dagger}%
\end{array}
\right)  \right\}  {\small 2n+2q,}\tag{61}\\
\overline{\eta}^{M}  &  =\overset{2m+2p}{\overbrace{\left(
\begin{array}
[c]{cc}%
b^{\dagger m}, & \eta^{\dagger l}%
\end{array}
\right)  }};\text{ }\eta_{M}=\left(
\begin{array}
[c]{c}%
b_{m}\\
\eta_{l}%
\end{array}
\right)  \tag{62}%
\end{align}
where we have defined%
\begin{align}
a_{c}^{\dagger}  &  =\frac{1}{\sqrt{2}}\left(  \lambda_{\alpha}+\overline{\mu
}^{\overset{\cdot}{\alpha}}\right) \tag{63}\\
b_{m}  &  =-\frac{1}{\sqrt{2}}\left(  \lambda_{\alpha}-\overline{\mu
}^{\overset{\cdot}{\alpha}}\right)  \tag{64}%
\end{align}

\section{Concluding remarks}

In this paper we construct the $4$-dimensional $N$-extended quaternionic
superspaces from the supersymmetric extension of the ordinary Penrose-twistor
formulation, with $N=p+k.\left(  p=N-k\right)  $

These quaternionic superspaces have $4+k\left(  N-k\right)  $ bosonic
quaternionic coordinates and $4N$ fermionic quaternionic coordinates where
each coordinate is a quaternion composed by four fields (bosons and fermions respectively).

The superspace coordinates are determined in terms of \textit{two}
quaternionic supertwistors corresponding to the fundamental Ferber's
representation and $k$ quaternionic fermionic supertwistors corresponding to
the Litov-Pervushin representation as we show in a previous section.

The biquaternionic construction for $N=8$ it is the more convenient to
represent the SM\ with $N=2k$, being also possible the nonlinear realization
of the symmetries. This fact is achieved due the (super) symplectic (almost
complex) structure of this construction conveniently extended to an even-
ortogonal group $O\left(  2N\right)  .$ The reason is fundamented by
Ambrose-Singer theorem and Extended Rothstein theorem(R-C-L theorem see
\cite{Alg} and ref. therein) that clearly relates the graded structure of the
tangent space and the field content of the realized physical theory e.g:
GUT\ containing the standard model..

Having into account the developments made here, in the next work
\cite{diegoVik} we will perform the nonlinear realization for the N=8 case to
obtain GR+SM.

\section*{Acknowledgment}

D.J. Cirilo-Lombardo is grateful to the Bogoliubov Laboratory of Theoretical
Physics-JINR\ and CONICET-ARGENTINA\ for financial support. DJC-L and VNP\ are
very grateful to A.B. Arbuzov for useful comments and insights.

\end{document}